\documentclass[12pt,preprint]{aastex}

\shorttitle{A tidal disruption event in RX\,J1242.6--1119A}
\shortauthors{St. Komossa et al.}
\begin{document}

\def\lsim{\mathrel{\lower .85ex\hbox{\rlap{$\sim$}\raise
.95ex\hbox{$<$} }}}
\def\gsim{\mathrel{\lower .80ex\hbox{\rlap{$\sim$}\raise
.90ex\hbox{$>$} }}}

\title{A Huge Drop in X-ray Luminosity of the Non-Active Galaxy
RX\,J1242.6--1119A, and First Post-Flare Spectrum -- Testing the Tidal
Disruption Scenario}

\author{Stefanie Komossa$^{1}$, Jules Halpern$^{2}$,
Norbert Schartel$^{3}$, G\"unther Hasinger$^{1}$,
Maria Santos-Lleo$^{3}$,
Peter Predehl$^{1}$}

\affil{$^{1}$Max-Planck-Institut f\"ur extraterrestrische Physik,
Giessenbachstrasse 1, 85748 Garching, Germany; skomossa@mpe.mpg.de,
ghasinger@mpe.mpg.de, prp@mpe.mpg.de\\
$^{2}$Department of Astronomy, Columbia University,
         550 West 120th Street, New York, NY 10027, USA;
jules@astro.columbia.edu\\
$^{3}$XMM-Newton Science Operation Center, European Space Agency,
         Villafranca del Castillo, Apartado 50727, E-28080 Madrid,
Spain;
        nscharte@xmm.vilspa.esa.es, msantos@xmm.vilspa.esa.es }

\begin{abstract}
In recent years, indirect evidence has emerged suggesting that many
nearby non-active galaxies
harbor quiescent supermassive black holes.
Knowledge of the frequency of occurrence of black holes,
of their masses and spins, is of 
broad relevance for studying black hole growth
and galaxy and AGN formation and evolution.
It has been suggested that
an unavoidable consequence of the existence of supermassive black holes,
and the
best diagnostic of their presence
in non-active galaxies, 
would be occasional tidal disruption
of stars captured by the black holes.
These events manifest themselves in form of luminous flares powered
by accretion of debris from the disrupted star into the black hole.

Candidate events among optically non-active 
galaxies emerged in the past few years.
For the first time, we have looked 
with high spatial and spectral resolution
at one of these most extreme variability events ever
recorded among galaxies.
Here, we report measuring a factor $\sim 200$ drop in luminosity
of the X-ray source RX\,J1242--1119 with the X-ray observatories
{\it Chandra} and {\it XMM-Newton}, and perform 
key tests of the favored outburst scenario,
tidal disruption of a star by a supermassive black hole.
We show that
the detected ``low-state'' emission has properties such that it must still
be related to the flare.
The power-law shaped post-flare X-ray spectrum
indicates
a ``hardening'' compared to outburst.
The inferred black hole mass, the amount of liberated energy,
and the duration of the event favor an accretion event
of the form expected from the (partial or complete) tidal
disruption of a star.
\end{abstract}

\keywords{galaxies: individual (RXJ1242.6--1119) --- galaxies: nuclei ---
                      X--rays: galaxies}

\section{Introduction}

The X-ray luminous nuclei of active galaxies (AGN) are believed to be
powered
by accretion of gas onto supermassive black holes.
There is now growing evidence
that many nearby {\em non-active} galaxies
harbor quiescent, weakly or non-accreting black holes  
(see reviews by Kormendy \& Gebhardt 2001; Richstone 2002).
Studies of the abundance of black holes, of
their masses and their spins, shed light on 
the mechanisms of black hole growth
and of galaxy and AGN formation and evolution.
Possibly the most direct means of detecting supermassive black holes
at the centers of galaxies, and
an unavoidable consequence of their existence,
would be occasional tidal disruption of stars and subsequent accretion
of their debris by these supermassive black holes (e.g., Hills 1975;
Gurzadyan \& Ozernoi 1979; Carter \& Luminet 1982; Rees 1988, 1990;
Wang \& Merritt 2004).
The events would appear as luminous flares of radiation
emitted when the stellar debris is accreted by the black hole.

Stellar capture and disruption is -- 
apart from accretion of gas and black hole merging --
one of the three major processes studied in the context
of black hole growth (e.g., Frank \& Rees 1976). 
The relative importance of these  three processes in feeding black holes
is still under investigation.
Zhao, Haehnelt, \& Rees (2002)
and Merritt \& Poon (2003) recently pointed out
that tidal capture may play an important role in explaining
the 
notable $M_{\bullet}-\sigma$
relation of galaxies.
Given the intense theoretical attention the topic of stellar tidal
disruption receives (see \S3 of Komossa 2002, and references
therein),
it is of great interest to see whether these events do occur in nature,
and how frequent they are.

Making use of the unique capability provided by the
All-Sky Survey (RASS, Voges et al. 1999) of the X-ray satellite {\it ROSAT\/},
giant X-ray flares from the directions of a few nearby galaxies
were discovered (Komossa \& Bade 1999; Komossa \& Greiner 1999; Grupe et
al. 1999;
Greiner et al. 2000). These flares were characterized by
huge peak X-ray luminosities reaching $\sim 10^{44}$ ergs~$s^{-1}$,
large amplitudes of decline, ultrasoft X-ray spectra,
and absence of Seyfert activity in ground-based optical spectra (see
Komossa 2002
for a review).
The target of the present study, RX\,J1242--1119,
was first detected by {\it ROSAT\/} in 1992 (Komossa \& Greiner 1999)
during a pointed observation with the Position Sensitive Proportional Counter (PSPC).
At that time it showed a very soft X-ray spectrum
with $kT_{\rm bb} \simeq 0.06$~keV,
and an X-ray luminosity of $L_{\rm x} \simeq 9 \times 10^{43}$ ergs~$s^{-1}$,
which is exceptionally large
given the absence of any signs of Seyfert activity of RX\,J1242--1119
in ground-based 
(Komossa \& Greiner 1999) and {\it Hubble Space Telescope} ({\it HST},
Gezari et al. 2003) optical spectra.
The association with one of the two, previously unstudied
non-active galaxies at redshift $z=0.05$ located
in the X-ray position error circle
(named RX\,J1242--1119A and RX\,J1242--1119B for lack of a better
designation) however, remained uncertain because 
of the large {\it ROSAT\/} positional uncertainty,
in this case 40$^{\prime\prime}$ (Figure 1).

Considering the extraordinary properties of this and a few similar
X-ray events, particularly their enormous luminosity output,
it is of utmost importance to understand what is happening in
these systems.
For the first time, we have now looked at one of these flare events
with high spatial and spectral resolution, in order to
confirm the counterpart,
follow the long-term temporal behavior, study the spectral
evolution and measure the post-flare spectrum, 
and to use these results to test the favored outburst model:
tidal disruption of a star by a supermassive black hole.
Among the few known X-ray flaring galaxies,
RX\,J1242--1119 was our target of choice
for follow-up X-ray observations
because it flared most recently, so the probability of catching the source in
the declining phase was highest. Also, because there were two galaxies in
the
{\it ROSAT\/} X-ray error box, the correct counterpart
needed to be confirmed.
A Hubble constant of 50 km~s$^{-1}$~Mpc$^{-1}$ is used throughout this paper.

\section{Data Reduction and Results}

\subsection{{\it Chandra}}
We observed the field of RX\,J1242--1119 with the
Advanced CCD Imaging
Spectrometer (ACIS-S) instrument
on-board the {\it Chandra X-ray Observatory} (Weisskopf et al. 2002)
for 4.5 ks on 2001 March 9.
The X-ray photons from the target source were
collected on the back-illuminated S3 chip of ACIS.

The {\it Chandra} data allow us to locate
precisely the counterpart of the X-ray source.
We find
that the X-ray emission peaks at coordinates
R.A.=$12^{\rm h}42^{\rm m}38.\!^{\rm s}55$,
decl.=$-11^{\rm o}19^{\prime}20.\!^{\prime\prime}8$ (equinox J2000).
Within the errors, this position is
coincident with the center of the optically brighter
of the two galaxies, RX\,J1242--1119A (Figure 1).
An offset in position of $\approx 1^{\prime\prime}$ in R.A.
can be traced to residual uncertainties in
the absolute pointing accuracy of the X-ray telescope.
Neither {\it Chandra} nor {\it XMM-Newton} detected any other
X-ray source within the {\it ROSAT\/} X-ray position error circle,
including from the optically fainter galaxy RX\,J1242--1119B.

We do not find evidence for significant source extent.
The photons detected by {\it Chandra} from RX\,J1242--1119A
fall within a radius of $1^{\prime\prime}$, and the radial profile is consistent
with a point source. This indicates that we are still seeing late-phase
flare-related emission rather than persistent, extended X-ray emission
originating from the host galaxy itself.  

A further  result, immediately obvious upon inspection of the
{\it Chandra} data, is the huge drop in X-ray flux of RX\,J1242-1119 
compared to its last observation by {\it ROSAT\/}.
Only 18 source photons were detected by {\it Chandra};
the implied amplitude of variability is quantified below.

\subsection{\it{XMM-Newton}}

Given the strong indications, based on the {\it Chandra} observation,
that the X-ray emission from RX\,J1242--1119A is still flare-dominated,
we asked for a target-of-opportunity observation with {\it XMM-Newton}
in order to obtain, for the first time,
a good-quality X-ray
spectrum of one of the flaring sources, and to follow the long-term
spectral evolution before the source had declined to
non-detectability.
While the strength of {\it Chandra} is its high spatial
resolution of better than $1^{\prime\prime}$, {\it XMM-Newton}
has higher throughput (the 18 source photons registered with
{\it Chandra} do not allow spectral fitting).

The {\it XMM-Newton} (Jansen et al. 2001) observation of RX\,J1242--1119
was performed on 2001 June 21-22, with a duration of 24.3 ks.
Data from the EPIC-pn, MOS1, and MOS2 detectors were used for
analysis.
The observation was first checked for flares in the background
radiation; none were detected.
Photons were then extracted from a circular area
of radius $0.\!^{\prime}3$ centered on the target source.
For the MOS cameras the background was determined in an
annulus around the source with an inner radius
of $1.\!^{\prime}0$ and an outer radius of $3.\!^{\prime}0$.
The background counts for the pn observation were selected
in a circular region close to the target source.
The total source count rate, measured
in the EPIC-pn detector, amounts to 0.0137 counts~s$^{-1}$
in the $0.3-5$~keV energy band.
No systematic decrease (or increase) of the count rate
is present during the observation.

The X-ray {\it spectrum} 
yields important information on the physics governing the post-flare
evolution.  X-ray emission from RX\,J1242--1119A
is detected in all three of the following energy
intervals: 0.3--0.75~keV, 0.75--1.5~keV, 1.5--5~keV,
with count rates of 0.0073, 0.0040, and 0.0024 counts~s$^{-1}$, respectively.
For spectral analysis, source photons were  
binned such that each spectral bin contains $>25$ photons.
The resulting spectrum is best described by a power law
with photon index $\Gamma_{\rm x} = -2.5 \pm 0.2$
($\chi^{2}_{\nu} = 1.1$; Figure 2). 
We do not find evidence for strong excess cold
absorption. The absorption towards RX\,J1242--1119A is
consistent with the Galactic value of $N_{\rm H} = 3.74 \times 10^{20}$ cm$^{-2}$.
Thermal (Raymond-Smith and MEKAL) X-ray
emission models yield significantly
worse spectral fits   
if the metal abundance of the X-ray emitting gas
is constrained not to fall significantly below 0.1 times the solar value.

Using the best-fitted power-law spectral model,
the {\it XMM-Newton} ``low-state'' luminosity of RX\,J1242--1119 is
$L_{\rm x} = 4.5 \times 10^{41}$ ergs s$^{-1}$ in the 0.1--2.4 keV band.
Compared to the peak luminosity of $9 \times 10^{43}$ ergs~s$^{-1}$
measured with {\it ROSAT\/},
this corresponds to a factor $\approx 200$ decline.

Finally, we used the Optical Monitor (OM) data to estimate
the $B$ and $U$ magnitudes of RX\,J1242--1119A (and B).
The OM observations were performed in standard image mode, with $U$, $B$,
and two broad-band UV filters.
Five high- and one low-resolution image
per filter were collected with an exposure time of
1000~s each. The data files were processed with {\it sas 5.4.1} and
with the task
{\it omichain} to perform required corrections for tracking, bad
pixels, fixed pattern, and flat-field.
Both galaxies are detected in the $U$, $B$, and $UVW1$ filter bands,
and their angular separation is large enough to avoid confusion.
Total source counts were measured for each galaxy in an aperture radius
of 12 pixels as recommended to account for the broad wings of the OM
point-spread function. Corrections for coincidence losses and dead-time
were applied to the total count rates, and then background was subtracted.
A mean corrected count rate
was then computed for each filter band. To estimate the magnitudes, the
most recent zero-point corrections (as of 2003 November) were applied.
This yields the following $B$ and $U$ magnitudes for RX\,J1242--1119A(B):
$m_B=17.56 \pm 0.05(18.72 \pm 0.06)$ mag and $m_U=17.80(18.83)$ mag.

\section{Discussion}

We have utilized the complimentary abilities of the 
X-ray observatories {\it Chandra} and {\it XMM-Newton}
to study the peculiar {\it ROSAT\/} source RX\,J1242--1119.
With {\it Chandra}, we found that the source of the
bright X-ray emission detected by {\it ROSAT\/} is the galaxy
RX\,J1242--1119A. The flare emission has declined dramatically by
a factor of $\approx 200$, but
has not yet faded away completely. 
{\it XMM-Newton} then provided us with the best post-flare spectrum
ever taken of any of
the few flaring, optically inactive galaxies. The spectrum  
is of power-law shape;
there is no evidence for excess absorption,
and the post-flare spectrum is harder than the flare maximum.

There is an additional argument that
the detected point-like low-state X-ray emission does not arise from 
the ISM of the host galaxy, but is related to the mechanism
that produced the flare itself.
Compared with the blue luminosity of RX\,J1242--1119A,
inferred from the extinction-corrected blue magnitude
$m_{B,0}=17.43$ mag measured with the OM,
the X-ray emission of this galaxy is still very high, even in its low state.
Its ratio of X-ray to blue luminosity, ${\rm log}\,(L_{\rm x}$/$L_B) \approx 31.6$,
is above the upper end observed so far for field early-type galaxies 
(Irwin \& Sarazin 1998; O'Sullivan et al. 2001).

While the X-ray observations presented here
demonstrate that the post-flare emission
is associated with the galaxy RX\,J1242--1119A,
it was established previously that
the optical spectrum of RX\,J1242--1119A is that of a non-active
galaxy (Komossa \& Greiner 1999, Gezari et al. 2003).
In addition to other arguments
presented earlier, the absence of any significant
excess X-ray absorption above the Galactic value
argues against the scenario of
an extremely X-ray variable active galactic nucleus
that is completely obscured optically.
We find that the galaxy RX\,J1242--1119A is not detected in
the NRAO VLA Sky Survey (NVSS, Condon et al. 1998) at 1.4~GHz,
also consistent with the absence of a {\em permanent\/} (hidden)
active nucleus.

The X-ray observations reported here
place important constraints on the total duration of
the flare maximum.
The source was not detected in 1990 in a short exposure
during the RASS.  It was ``on'' in 1992,
and had dropped
dramatically in flux during the {\it Chandra} and {\it XMM-Newton}
observations in 2001,
limiting the ``on-state'' to $\leq 10$~yr.
The high outburst luminosity and amplitude of variability
strongly suggest that a supermassive
black hole at the center of RX\,J1242--1119A 
is the ultimate power source.
The new observations presented here
are consistent with and corroborate
the previously favored 
tidal disruption scenario to explain this flare event.

Independent of the X-ray properties of RX\,J1242--1119A,
its blue magnitude measured by the OM
can be used to estimate the
mass of the black hole at the center of the galaxy.
The correlation between the absolute
blue magnitude of the bulge of an elliptical galaxy
and the mass of the central black hole
(Ferrarese \& Merritt 2000)
predicts a black hole mass of $\approx 2\times 10^8\,M_{\odot}$.
This estimate is uncertain by a factor of several, given the
scatter in the relation between bulge luminosity and black
hole mass (e.g., Ferrarese \& Merritt 2000, McLure \& Dunlop 2002).
The observations are consistent with either partial disruption
of a giant star, or complete disruption of a 
solar-type star{\footnote{the
latter only if the tidal radius
is located outside the Schwarzschild radius, i.e.,
for $M_{\rm BH} \lsim 10^8\,M_{\sun}$ (e.g., Hills 1975); or 
for larger black hole masses, if the black hole is of Kerr type
and the star approaches from a favorable direction (Beloborodov et al. 1992)}},
where either only part of the debris from the solar-type star is accreted
while the rest becomes unbound, or where part of the debris is accreted
in a radiatively inefficient mode. 

To estimate (a lower limit on) the amount of stellar debris accreted
by the black hole, we forced the high-state and low-state luminosities
to follow a $t^{-5/3}$ law expected for the ``fall-back'' phase
(e.g., Rees 1990; Li et al. 2002) of tidal disruption. The integral
$\int L(t)\,dt$ then gives the total emitted energy,
$E \simeq 1.6 \times 10^{51}$~ergs, where we started the
integration at the time the source was first observed in the high-state
by {\it ROSAT\/}.
This requires accretion of $M \simeq 0.01\,\eta_{0.1}^{-1}\,M_{\odot}$
of stellar material, where $\eta=0.1\,\eta_{0.1}$ is the efficiency.

Although the X-ray emission of RX\,J1242--1119A dropped substantially,
does this necessarily mean that the {\it total} accretion luminosity declined,
or is it possible that the emission just shifted out of the X-ray band?
Part of the flux may have softened into the unobservable
EUV part of the spectrum, as expected in some tidal disruption models
(e.g., Cannizzo et al. 1990),
but it is more likely that the total luminosity has declined since
outburst.  Otherwise, we might expect to see optical emission lines
excited by the hypothesized bright EUV continuum; such emission lines
do not appear in {\it HST\/} spectra of RX\,J1242--1119A taken several
years after the outburst (Gezari et al. 2003).

With all observational tests possible now performed,
more detailed comparison with theory has to await refined model
calculations of stellar disruption.

\section{Concluding Remarks}

The results presented here demonstrate the effectiveness of
combining the abilities of the two most powerful X-ray observatories
in orbit to study one of the most
extreme variability events ever recorded among galaxies.
In our interpretation of these observations, we are seeing the
post-disruption phase of a close encounter of a star with a
central supermassive black hole, into which some of
the tidal debris is accreting.
Continued X-ray monitoring of RX\,J1242--1119A will enable us to follow
the expected further decline in luminosity. 

Future X-ray (all)-sky surveys planned for missions like
{\it DUO\/}, {\it ROSITA\/}, {\it LOBSTER\/},
and {\it MAXI\/} will be very useful in finding
new flare events, while their detailed study
will become possible with future high-throughput
X-ray missions like {\it XEUS\/} and {\it Con-X\/}.
We may then be able to probe the regime
of strong gravity, since the temporal evolution
of the flare X-ray emission is expected to depend
on relativistic precession effects in the Kerr metric.

The results presented here will impact further model
calculations of the tidal disruption process,
which are still complex and challenging,
and are expected to motivate new studies of the flare host galaxies as
well as an expanded search
for similar flare events in existing and future
{\it Chandra} and {\it XMM-Newton} archives, including the deep fields.

Such observations of flares open up a new window to study supermassive
black holes and their environment, and the physics of accretion events
in otherwise inactive galaxy nuclei.

\acknowledgments
This work is based on observations obtained with
{\it XMM-Newton}, an ESA science
 mission with  instruments and contributions directly funded by
ESA Member States and NASA, and on observations obtained with the NASA
mission {\it Chandra}. 
In Germany, the {\it XMM-Newton} project is supported by the
Bundesministerium f\"ur Bildung und Forschung/Deutsches Zentrum f\"ur
Luft- und Raumfahrt, the Max-Planck Society and the
Heidenhain-Stiftung.
We thank the {\it XMM-Newton} and {\it Chandra} teams for
their help which made these observations
possible,
and Peter Edmonds, Joachim Tr\"umper
and an anonymous referee for their useful comments.  
I, St.K. gratefully remember Hartmut Schulz for ongoing discussions on
many topics in astrophysics including the flares, for sharing his broad
knowledge
and deep insight into physics, for patiently listening when I needed
somebody to talk to, for intense collaborations on NGC\,6240 ever since
I started in astrophysics, and for sharing his enthusiasm for NGC\,4151.
Hartmut passed away in August 2003.



\begin{figure}
\begin{center}
\begin{minipage}[t]{14cm}
\includegraphics[width=11.0cm, angle=0, clip=]{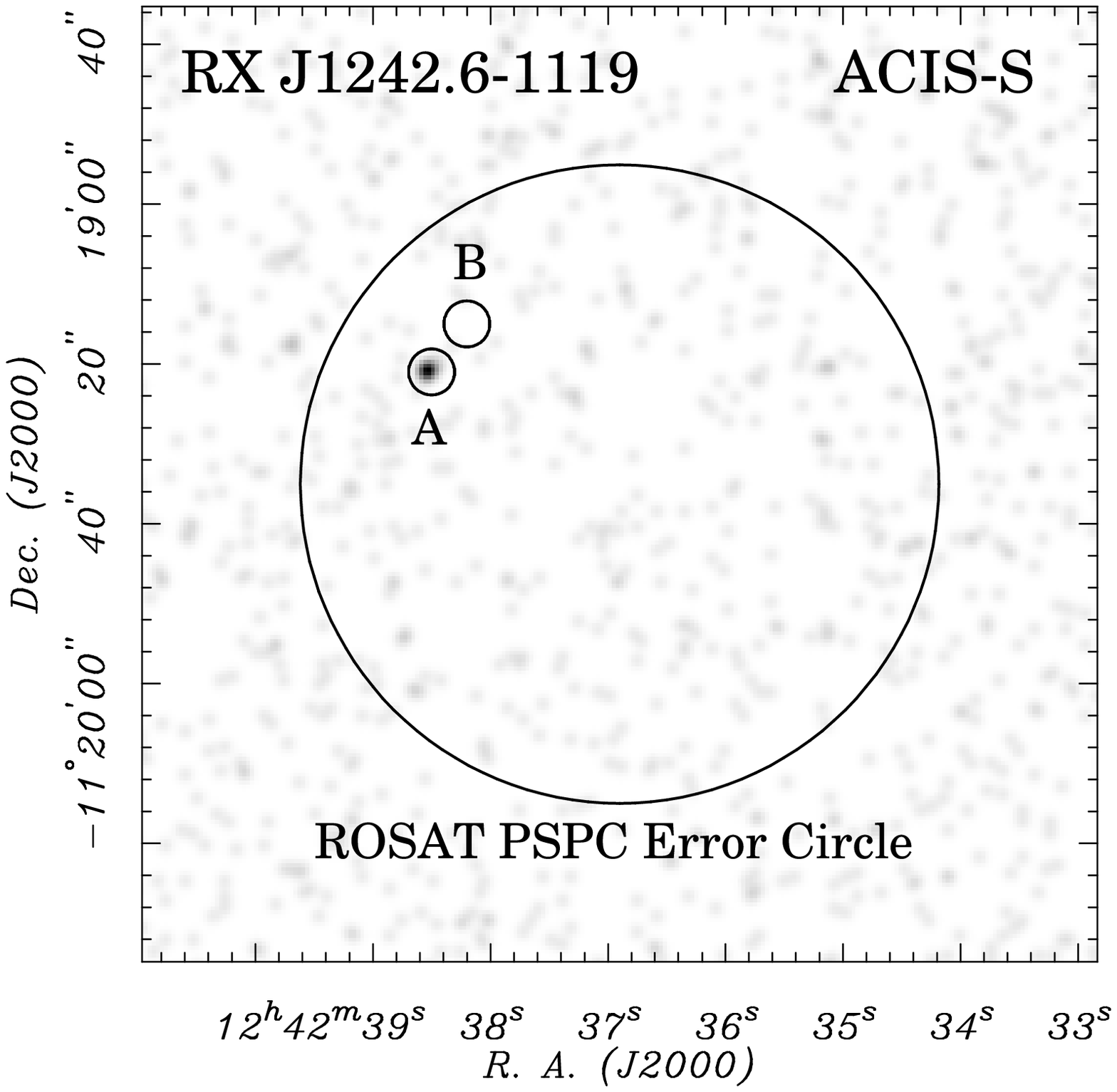}
\end{minipage}
\end{center}
\caption{{\it{Chandra}} X-ray image of the area around
RX\,J1242--1119A. The {\it large circle} corresponds to
the {\it ROSAT\/} error box, while the two
{\it small circles} mark the
optical positions of the two galaxies RX\,J1242--1119A and B.  
}
 \end{figure}

\begin{figure}
\begin{center}
\begin{minipage}[t]{14cm}
\includegraphics[width=9.0cm, angle=-90, clip=]{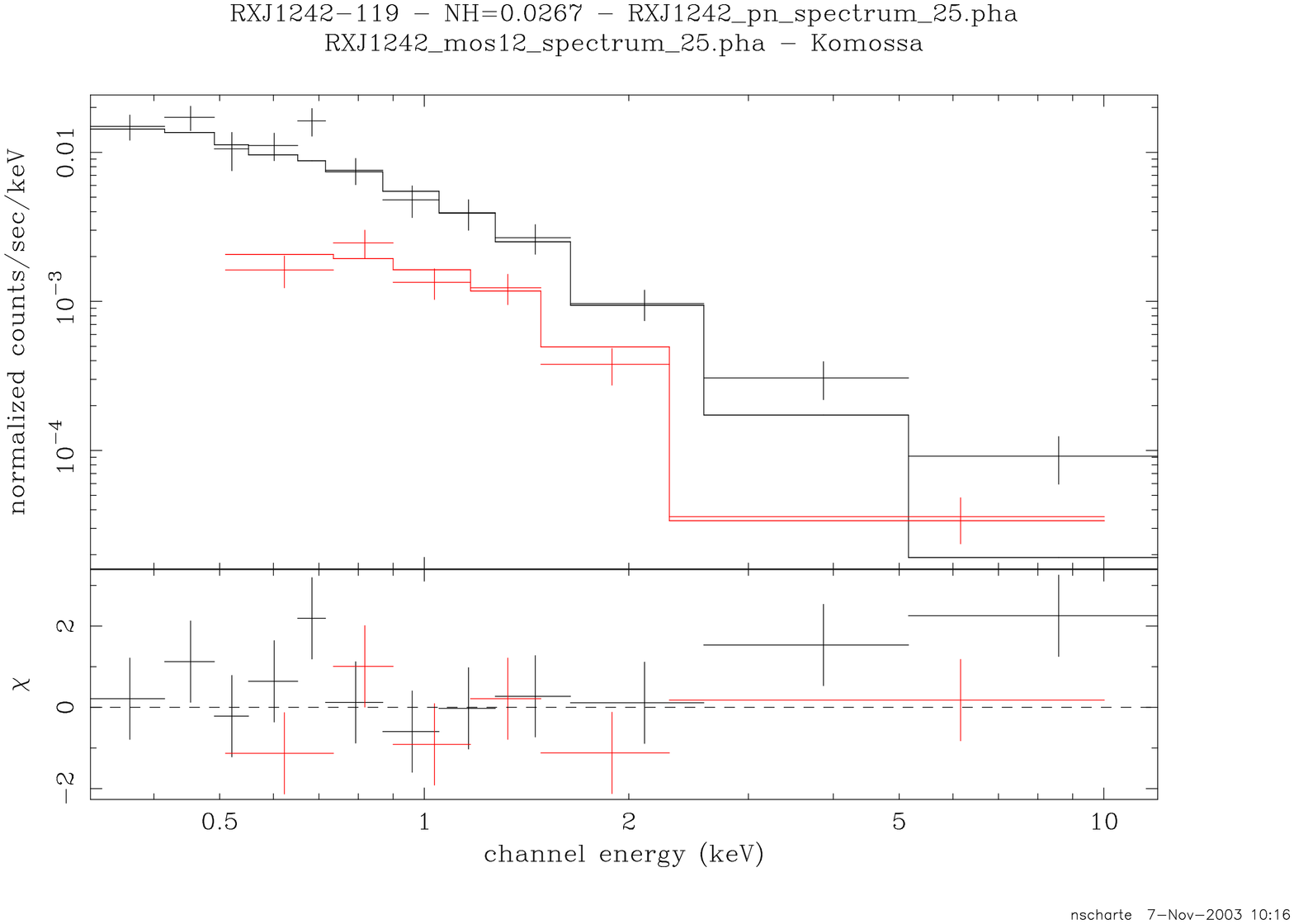}
\end{minipage}
\end{center}
\caption{{\it XMM-Newton} X-ray spectrum of RX\,J1242--1119,
and best-fitted power law model ({\it solid lines}).
{\it Upper symbols}: EPIC pn data.
{\it Lower symbols}: MOS 1 and 2 data binned together.
}
 \end{figure}

\end{document}